\documentclass[usegraphicx,usenatbib,useapjfonts,apjl]{emulateapj}

\def\plotone#1{\centering \leavevmode
\epsfxsize=\columnwidth \epsfbox{#1}}

\begin{document}

\title[Detectability of the effect of Inflationary non-Gaussianity on halo bias]{Detectability of the  effect of inflationary non-Gaussianity on halo bias}

\author{Licia Verde\altaffilmark{1,2} \&  Sabino Matarrese \altaffilmark{3}}
\altaffiltext{1}{ICREA:  Instituci\'o  Catalana de Recerca i Estudis Avancat \& Instituto de Ciencias del Cosmos (ICC) Universidad de Barcelona, Marti i Franques 1, 08028 Barcelona, Spain; liciaverde@icc.ub.edu}
\altaffiltext{2}{Theory Group, Physics Department, CERN, CH-1211, Geneva 23, Switzerland}
\altaffiltext{3}{Dipartimento di Fisica ``G. Galilei", Universit\`a 
degli Studi di Padova and INFN, Sezione di Padova, via Marzolo 8, 35131, Padova, Italy; sabino.matarrese@pd.infn.it}

\begin{abstract}
We consider the description of the clustering of halos for physically--motivated types of  non-Gaussian initial conditions. 
In particular, we include non-Gaussianity of the type arising from single-field slow-roll, multi fields, curvaton (local type), higher-order 
derivative-type (equilateral), vacuum-state modifications (enfolded-type)  and horizon-scale GR corrections type. 
We show that large-scale halo bias  is  a very sensitive  tool to probe non-Gaussianity, potentially leading, for some planned surveys, to a 
detection of non-Gaussianity arising from horizon-scale GR corrections.  In tandem  with cosmic microwave background constraints, the halo bias approach can help enormously to discriminate  among different  shapes of non-Gaussianity  and thus among models for the origin of cosmological perturbations.
\end{abstract}


\section{Introduction}
A powerful test of the generation mechanism for cosmological perturbations in the early universe is offered by constraining non-Gaussianity of the primordial fluctuations.  
The leading theory for the origin of primordial perturbations is inflation: non-Gaussianity is a sensitive probe of aspects 
of inflation that are difficult to probe otherwise, namely  the 
interactions of the field(s) driving inflation. 
While standard single-field models of slow-roll inflation lead to small departures from Gaussianity, non-standard scenarios  allow for a larger level of non-Gaussianity (\citet{BKMR04} and references therein). 
In particular, large non-Gaussianity can be produced if any of the conditions below is violated:
{\it i)} single field, {\it ii)} canonical kinetic energy {\it iii)} slow roll and {\it iv)} adiabatic (Bunch-Davies) initial vacuum state.
The standard observables to constrain non-Gaussianity are the cosmic microwave background (CMB) and large-scale structure and in particular their bispectrum (or three-point correlation function), 
although halo abundance and clustering can offer complementary  constraints.  
It has  recently been shown \citep{linde/mukhanov:1997,lyth/ungarelli/wands:2003, BabichCreminelli04,chen/etal:2007,holman/tolley:2008,chen/easther/lim:2007,langlois/etal:2008, Meerburgetal09} 
that violation of each  of the conditions above produce its own signature in the bispectrum shape (i.e. the dependence on the form of the triangle made by its three wave vectors).

The type of non-Gaussianity arising in standard inflation is of the type  \citep{SalopekBond90, Ganguietal94,VWHK00, KS01}
\begin{equation}
\Phi=\phi+f_{\rm NL}\left(\phi^2-\langle \phi^2 \rangle\right)\,,
\label{eq:fnl}
\end{equation}
where $\Phi$ denotes Bardeen's gauge-invariant potential, which,   
on sub-Hubble scales reduces to the usual Newtonian peculiar 
gravitational potential, up to a minus sign and $\phi$ denotes a Gaussian random field. The non-Gaussianity parameter $f_{\rm NL}$ is often considered to be constant, in which case this is called local non-Gaussianity  and its bispectrum is maximized for squeezed configurations (where one wave vector is much smaller than the other two).  Non-Gaussianity  of the local type is generated in  
standard inflation (in this case $f_{\rm NL}$ is expected to be of  the same order of the slow-roll parameters)  and for multi-field models.
Note, however, that an expression like Eq.(\ref{eq:fnl}) is not general and there are many inflationary models which predict different types of deviations from Gaussianity. In general, their non-Gaussianity is specified by their bispectrum. There are some cases  where the trispectrum may be important (when, for example, the bispectrum is zero) but in general one expects the trispectrum contribution to be sub-dominant compared to the bispectrum one. 

While CMB and large-scale structure can measure the bispectrum shape-dependence and thus can in principle discriminate the shape of non-Gaussianity (e.g., \cite{FergussonShellard08}  and references therein) there are  also other powerful probes.
 One technique is based on the abundance  \citep{RB00, RGS00,MVJ00, VJKM01, Loverdeetal07,KVJ09,JV09}  of  rare events such as  dark matter density peaks  as they trace the tail of the underlying distribution. This probe is sensitive to the  primordial skewness: being the skewness an integral over all bispectrum shapes, this probe cannot  easily discriminate among different shapes of non-Gaussianity.  

Recently, \cite{DDHS07} and \cite{MV08}(hereafter MV08) have shown that primordial non-Gaussianity affects the clustering of dark matter halos inducing (in the case of local non-Gaussianty) a scale-dependent bias on large scales. This effect, which goes under the name of non-Gaussian halo bias, is particularly promising, yielding already stringent constraints from existing data  \citep{Slosaretal08, AfshordiTolley08} and, with forthcoming data,  offers  the potential  to rival the constraints achievable from an ideal CMB survey \citep{CVM08}. 
Despite being so promising, this effect has not been extensively considered for non-Gaussianity  which is not of the local type. In fact, the derivation of \citet{DDHS07} and \cite{Slosaretal08} can only deal with  local non-Gaussianity. \cite{Taruyaetal08} use a perturbation theory approach and obtain expressions for the local and equilateral type of non-Gaussianity.

On the other hand, the approach of MV08 is general enough to yield an expression for the large-scale non-Gaussian  halo bias for any type of non-Gaussianity specified by its bispectrum. 
Here we explore the effect on the clustering of halos of  physically motivated primordial non-Gaussianity different from the local  case. We briefly review the local case, examine equilateral  and  enfolded types. We then concentrate on a type of   non-Gaussianity  sourced by inflation  and affecting large-scale structures, only arising when considering general-relativistic  (GR) corrections to  the standard Newtonian treatment.
The paper is organized as follows: in \S 2  we briefly  review the \cite{MV08} description of the non-Gaussian halo bias. In \S 3, we consider physically motivated primordial non-Gaussianity different from the local  case and introduce the non-Gaussianty arising from GR corrections.  In \S 4, we consider the constraints that planned experiments can place on these non-Gaussianities and present our  conclusions.

\section{Non-Gaussian halo bias for a given primordial  bispectrum}
Halo clustering can be modeled by assuming that halos correspond to regions where the (smoothed) linear dark matter density field exceeds a suitable threshold. For massive halos, the threshold is high compared to the field {\it rms}.
The MV08 approach  to non-Gaussian halo bias relies on the fact that the two-point correlation function of regions  above a high threshold for a general non-Gaussian field has an analytical expression \citep{MLB86} which depends on all higher-order (connected) correlations.  For most inflationary models the expression can be truncated so that it includes only terms up to  the three-point correlation function.
With the additional assumptions of  small non-Gaussianity  and separations that are much larger than the  Lagrangian halo radius, MV08 obtain  that the halo power spectrum $P_h(k,z)$ is related to the dark matter density field  $P(k,z)$ by:
\begin{equation}
P_h(k,z)=\frac{\delta^2_c(z)P(k,z)}{\sigma_M^2 D^2(z)}[1+2\delta_c(z)\beta(k)]\,.
\end{equation}
Here,  $\delta_c$ is the critical collapse threshold: $\delta_c(z)=\Delta_c(z)/D(z)$, where $\Delta_c(z)$ is the linearly extrapolated over-density 
for spherical collapse; it is $1.686$ in the Einstein-de Sitter case, 
while it slightly depends on redshift for more general 
cosmologies. $D(z)$ is the linear growth factor, which depends on the background cosmology; $\sigma_M \equiv \sigma_R$ is the {\it rms} value of the underlying (linear) dark matter fluctuation field at $z=0$, smoothed on a scale $R$ related to $M$ by $M=\Omega_{m,0}3H_0^2/(8\pi G)(4/3)\pi R^3$, with $\Omega_{m,0}$ denoting the present-day matter density parameter; $H_0$ is  the present-day Hubble parameter, and $G$ is  Newton's constant. 
The effect of non-Gaussianity is enclosed in the function $\beta(k)$ which in this approach is: 
\begin{eqnarray}\label{eq:beta}
\beta(k)&=& \frac{{\cal G}}{8\pi^2\sigma_R^2{\cal M}_R(k)}\int dk_1 k_1^2 {\cal M}_R(k_1) \times \nonumber \\
&&\!\int_{-1}^1\!d\mu
{\cal M}_R\left(\sqrt{\alpha}\right)\frac{B_{\Phi}(k_1,\sqrt{\alpha},k)}{P_{\Phi}(k)}.
\end{eqnarray}
Here $B_{\Phi}$ denotes the expression for the primordial bispectrum of the Bardeen potential $\Phi$, $P_{\Phi}$ its power-spectrum. Since in Eq.~(\ref{eq:fnl}) $\Phi$ is the primordial potential deep in the matter dominated era, ${\cal G}\sim 1.3$ accounts for the fact that  the potential evolves in redshift  in a non-Einstein-de Sitter Universe; ${\cal M}_{R}(k)$ is related to the Poisson equation via:
\begin{equation}
\delta_{R}({\bf k})=\frac{2}{3} \frac{T(k) k^2}{H_0^2 
\Omega_{m,0}}W_R(k)\Phi({\bf k})\equiv {\cal M}_R(k)\Phi({\bf k}) \;,
\label{eq:defM}
\end{equation}
with $W_{R}(k)$ being the Fourier transform of the top-hat window function of radius $R$.

Thus the correction to the standard halo bias due to the presence of primordial non-Gaussianity is:
\begin{equation}\label{eq:dboverb}
\frac{\Delta b_h}{b_h}= \frac{\Delta_c}{D(z)}\beta(k)\,.
\end{equation}
It is clear that a scale-dependent $\beta(k)$ could in principle give a distinctive detectable signature on the observed power spectrum. 

The above expressions were derived under the assumption that non-Gaussianity is a ``small" correction to the dominant Gaussian component of the primordial perturbations. An extensive discussion of the limits of this approximation and possible improvement is reported in \cite{CVM08}; here it will suffice to say that, looking for example at Eq. \ref{eq:fnl}, $f_{\rm NL}\phi^2\ll \phi$ and since $\phi={\cal O}(10^{-5})$, even $f_{\rm NL}$ of order $10^3$ (value which is already observationally excluded anyway) can be considered ``small''. 

\section{Inflationary non-Gaussianities}
In the local non-Gaussian case the bispectrum of the potential is given by:
\begin{equation}
B_{\Phi}(k_1, k_2, k_3)=2 f_{\rm NL}^{\rm loc} F^{\rm loc}(k_1,k_2,k_3)
\end{equation}
where
\begin{equation}
F^{\rm eq}(k_1,k_2,k_3)= P_{\Phi}(k_1)P_{\Phi}(k_2)+2 cyc 
\end{equation}

Looking at Eq.~(\ref{eq:beta}), it is easy to  write analytically the form for $\beta(k)$, and to compute it for the local non-Gaussian case:
\begin{eqnarray}
\beta^{\rm loc}(k)&=& \frac{2 f_{NL}^{\rm loc}}{{\cal M}_R(k)}\frac{\cal G}{8\pi^2\sigma_R^2}\int dk_1 k_1^2 {\cal M}_R(k_1)
P_{\Phi}(k_1)  \nonumber \\
&\times &\!\int_{-1}^1\!d\mu
{\cal M}_R\left(\sqrt{\alpha}\right)\left[ \!\frac{P_{\Phi}\left(\sqrt{\alpha}\right)}{P_{\Phi}(k)} 
+ 2 \right] . 
\end{eqnarray}

However, in the standard slow roll inflation $f_{\rm NL}^{\rm loc}$ is expected to be unmeasurably small. 
Inflationary models that  can  produce larger non-Gaussianity of the local form are those where the fluctuations of an additional light field, different from the  inflaton, contribute to the curvature perturbations we observe (see, e.g., \cite{BabichCreminelli04}), for example, curvaton models (e.g., \cite{Sasakicurvaton06, curvatonNG} and references therein) and multi-field models (e.g., \cite{BMRmulti,multifieldNG}).

On the other hand, inflationary models  with higher-derivative operators of the inflaton, such as, for example, the DBI model have a different type of non-Gaussianity, whose bispectrum is maximized for $k$ modes of similar scales (equilateral type; \cite{seery05,chen/etal:2007}).
The equilateral type of non-Gaussianity can be well described by the following template \citep{CreminelliNicolisSenatore05}:
\begin{equation}
B_{\Phi}(k_1, k_2, k_3)=6 f_{\rm NL}^{\rm eq} F^{\rm eq}(k_1,k_2,k_3)\,,
\end{equation}
where 
\begin{eqnarray}
F^{\rm eq}(k_1,k_2,k_3)&=& -P_{\Phi}(k_1)P_{\Phi}(k_2)+2 cyc \\ \nonumber
&-&2[P_{\Phi}(k_1)P_{\Phi}(k_2)P_{\Phi}(k_3)]^{2/3}\\ \nonumber
&+&( P_{\Phi}^{1/3}(k_1)P_{\Phi}^{2/3}(k_2)P_{\Phi}(k_3) +5 cyc.)\,.
\end{eqnarray}

General deviations from the simplest slow roll inflationary models are likely to have bispectra that are not well described by the two cases above (e.g., \cite{FergussonShellard08}).

Predictions for the primordial bispectrum evaluated in the regular Bunch-Davies vacuum state, are of  local or equilateral type, depending on  whether higher-derivative corrections play a significant role in the inflationary evolution. Non-Gaussianity generated by dropping the assumption that the vacuum is Bunch-Davies (modified initial state non-Gaussianity) is instead maximal for ``enfolded" (or ``squashed") configurations \citep{chen/etal:2007,holman/tolley:2008, Meerburgetal09}. The associated bispectrum is a complicated function  of the $k$'s which is not easily factorizable, but \cite{Meerburgetal09} proposed a  factorized enfolded template, which captures  very well the fetures of the modified initial-state bispectrum.
This enfolded factorizable  template is given by: 
\begin{equation}
B_{\Phi}(k_1, k_2, k_3)=6 f_{\rm NL}^{\rm enf} F^{\rm enf}(k_1,k_2,k_3)\,,
\end{equation}
where 
\begin{eqnarray}
F^{\rm enf}(k_1,k_2,k_3)&=&P_{\Phi}(k_1)P_{\Phi}(k_2)+2 cyc \\ \nonumber
&+&3[P_{\Phi}(k_1)P(k_2)P_{\Phi}(k_3)]^{2/3}\\ \nonumber
&-& (P_{\Phi}^{1/3}(k_1)P_{\Phi}^{2/3}(k_2)P_{\Phi}(k_3) +5 cyc.)\,.
\end{eqnarray}

While so far we have concentrated on non-Gaussianity generated during or right at the end of inflation, there are other non-Gaussian signals generated after the end of inflation and still before matter domination.
One source of non-Gaussianity that has received renewed attention is the second-order evolution of perturbations from inflation to matter domination  \citep{Bartolo05_1, Pillepich, BMR07,Fitzpatrick}. 
Large-scale structure probe clustering on scales that entered the horizon when radiation was important;  for large-scale modes that enter the horizon deep in matter dominated era, no perturbation growth is expected before matter domination so their entire growth history can be modeled in the standard way assuming matter domination. 
Non-Gaussianity induced by the non-linear growth of perturbations during radiation dominance, as discussed in \citet{BMR07} and \citet{Fitzpatrick}, does not affect the non-Gaussian halo-bias, as the relevant scales are different ($ > 100Mpc/h$ for the halo-bias).

On such large scales, there is, however,  an additional source of non-Gaussianity, arising from GR corrections on scales comparable to the Hubble radius.
This effect was first pointed out by  \citet{Bartolo05_1}, who performed the calculation both in the comoving and Poisson gauges. \citet{Fitzpatrick}  then extended the calculation to scales that entered the horizon during radiation dominance  and recovered the \cite{Bartolo05_1} expressions on large scales (those of interest here).
A relevant discussion on the issue can be found in \citet{WandsSlosar09} and \citet{Yoo09}. In particular, \citet{WandsSlosar09} argue that computing the density fluctuation field in the comoving time-orthogonal gauge yields the physical expression  needed to compute the large-scale halo bias. 

Using  the comoving density perturbation at second-order in the usual Poisson equation, allows us to obtain the GR correction to the primordial $f_{\rm NL}$. Indeed, one can easily check 
\citep{Bartolo05_1} that the resulting expression is unaffected by the presence of constant gauge modes, both at the linear and second-order levels.  

With this assumption, we obtain the following expression for the  large-scale structure  linear-regime bispectrum including GR-corrections: 
\begin{eqnarray}\label{eq.inflbisp1}
\!\!B_{\Phi}(k_1,k_2,k_3)&=& 2 \left [\frac{5}{3}(a_{\rm NL}-1) + f_{\rm NL}^{\rm infl,GR}(k_1,k_2, k_3)\right ] \nonumber  \\
&\times& P(k_1)P(k_2) + cyc. 
\end{eqnarray}
where $cyc.$ denotes terms with $\{k_2,k_3,k_1\}$ and $\{k_1,k_3,k_2\}$ and 
\begin{equation}
f_{\rm NL}^{\rm infl,GR}(k_i,k_j, k_k)=-\frac{5}{3}\left[ 1- \frac{5}{2}\frac{k_i k_j cos\theta_{ij}}{k_k^2}\right]
\label{eq:fnlinfl}
\end{equation}
with  $\theta_{ij}$ denoting the angle between the vectors $k_i$ and $k_j$. Eq. \ref{eq:fnlinfl} is obtained from Eq. 7 of \cite{Bartolo05_1} considering the terms that multiply the linear growth factor (thus dropping the Newtonian, gravitational instability terms), setting $a_{\rm nl}=1$, and translating to  the gravitational potential as usual. 

The first term on the RHS of Eq. \ref{eq.inflbisp1} is the primordial contribution from standard slow-roll inflation: $|a_{NL}-1|\ll 1$, being of the order of the slow-roll parameters \citep{Ganguietal94,acquavivaetal03, maldacena03}. 
All that follows is  the additional effect of interest here. 
Let us reflect on the meaning of this contribution,  which, as Eq. \ref{eq.inflbisp1} shows and because it is a second-order term like the $|a_{\rm NL} -1|$ one,  adds up to  the ``intrinsic" non-Gaussianity: 
\begin{itemize}
\item The RHS of Eq.~(\ref{eq.inflbisp1}) is non-zero even if  the --strictly speaking-- primordial contribution is zero. Inflationary models different from the standard slow roll would yield a different expression for the  first term in the RHS as briefly discussed above. 
\item The shape of $f_{\rm NL}^{infl,GR}$ is peculiar to the inflationary initial conditions in two aspects: {\it  i)} perturbations on super-Hubble scales are needed in order to initially feed the GR correction terms. In this respect, the significance of this term is analogous to the well-known large-scale anti-correlation between CMB temperature and E-mode polarization:  it is a consequence of the properties of the inflationary mechanism to lay down the primordial perturbations. {\it ii)} Initial conditions arising from standard slow-roll single field inflation imply that the second-order comoving curvature perturbation, defined as in \cite{SalopekBond90},  $\zeta^{(2}) \approx 0$, or equivalently  $a_{\rm NL} -1 \approx 0$: it is this very fact which leads to  the dominance of the GR corrections.
\end{itemize}

We note that in Eq.~(31) of  \citet{Pillepich}\footnote{Note that the Pillepich et al paper is about the bispectrum of the redshifted 21-cm fluctuations from the dark ages:  this bispectrum is present  in different large scale structure  tracers.} one can find the same quantity calculated in the Poisson gauge. We have numerically verified that using the Poisson-gauge expression the results for the non-Gaussian halo bias are practically unchanged 
(once the constant, pure gauge, modes appearing in the Poisson-gauge expression are ignored). This is indeed encouraging, as we recover that measurable quantities are  gauge-independent. 

Equation (\ref{eq:beta}) enables us to compute immediately the effect of these four different types of non-Gaussianity on the large-scale clustering of halos \footnote{Up to small, sub-dominant corrections involving details of  the transformation between Lagrangian and Eulerian bias factors.}. This is the necessary step to be able to quantify their detectability for this observable.  
\begin{figure}
\plotone{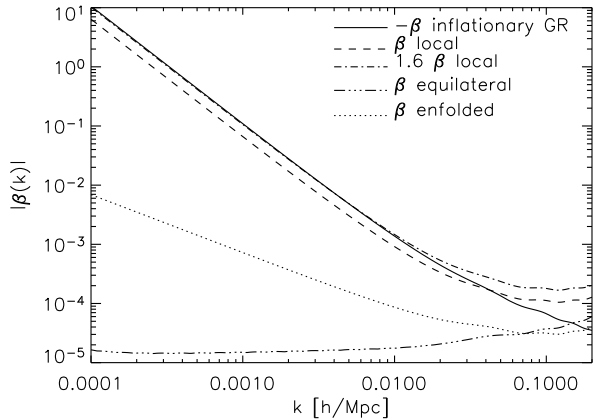}
\caption{The scale-dependence of the large-scale halo bias induced by non-zero bispectrum, indicated  by the $\beta$ function of Eqs. 3 for the four types of non-Gaussianity discussed in the text. The solid line shows the absolute value  of $\beta$ for the inflationary, GR correction large-scale structure bispectrum. Note that the quantity is actually  negative. The  dashed line shows $\beta$ for the local type of primordial non-Gaussianity for $f_{\rm NL}^{\rm loc}=1$ (the quantity is positive). It is clear that  the  scale-dependent bias effect due to the inflationary bispectrum mimics a local primordial non-Gaussianity with  effective $f_{\rm NL}$ $\sim  -1$ at $k>0.02$ $H$/Mpc and $\sim  -1.6$ for $k<0.01$ $h$/Mpc. The dot-dot-dot-dashed line shows the effect of equilateral non-Gaussianity for $f_{\rm NL}^{\rm eq}=1$ and the dotted line shows the  enfolded-type with $f_{\rm NL}^{\rm enf}=1$.}
\label{fig:beta}
\end{figure}

In Figure \ref{fig:beta} we show the scale dependence of the halo bias (i.e. the  $\beta(k)$ of eq. \ref{eq:dboverb}) generated by the  ``inflationary", GR-correction, large-scale structure bispectrum and we compare it with the one for the local-type primordial non-Gaussianity with $f_{\rm NL}^{\rm loc}=1$. On the scales of interest $\beta(k)^{\rm infl, GR} \propto \beta(k)^{\rm local}$ where the proportionality is a factor between $-1$ and $-1.6$. In other words the scale-dependent  bias effect of the inflationary, GR-correction,   non-Gaussianity is that of a local non-Gaussianity with $f_{\rm NL}\simeq -1.6$.

The dot-dot-dot dashed line shows the effect of equilateral type of non-Gaussianity for $f_{\rm NL}^{\rm eq}=1$, in agreement with the findings of  \cite{Taruyaetal08}. The effect for the $f_{\rm NL}^{\rm enf}=1$ enfolded-type (which closely describes effects of modified initial state)  is shown by the dotted line.

\section{Forecasts and Conclusions}

It is clear from Fig. \ref{fig:beta} that while  local and inflationary non-Gaussianity leave a strongly scale-dependent signature on the halo clustering, equilateral and enfolded type  have a much smaller effect.
This is not unsurprising: local and inflationary-type primordial non-Gaussianity have  strong mode-correlations between small and 
large-scale Fourier modes. For this reason,  biasing,  a small-scale phenomenon, can affect the power spectrum on very large scales. Equilateral and enfolded-type of non-Gaussianity have correlations for modes that are of comparable scales.

Following the calculations presented in \cite{CVM08}\footnote{\cite{CVM08} used the large-scale structure $f_{\rm NL}$normalization, while here we use the CMB normalization. See Table 1 of \cite{Grossietal09} for  the CMB-normalized forecasts.}  (taking into account the calibration on N-body simulations presented in \cite{Grossietal09})  we  can forecast what constraints could be achieved  from future surveys.

Equilateral  type of non-Gaussianity cannot   effectively be constrained using the scale-dependent halo-bias effect: the bispectrum of the CMB temperature fluctuations or of the galaxy distribution will be a much more powerful tool in this case. 
On the other hand, large-scale  halo-bias is  extremely promising for local and GR types.
Forecasts for the local type were presented in \cite{CVM08}, here we report that  a survey of the type of Euclid\footnote{http://sci.esa.int/science-e/www/area/index.cfm?fareaid=102}  can constrain GR corrections-type of  non-Gaussianity at the $1-\sigma$ level, while a survey like LSST\footnote{http://www.lsst.org}   could detect this signal at the  $2.2 - \sigma$ level. For the enfolded type we obtain  that a survey of the type of Euclid can yield a $1-\sigma$ error of $\Delta f_{\rm NL}^{\rm enf}=39$  while $\Delta f_{\rm NL}^{\rm enf}=18$ for LSST.

These error-bars are dominated by cosmic variance and could thus, in principle, be  reduced further  by a factor of $\sim$few using the approach proposed  by \cite{seljakCV}.
This opens up the possibility of detectng the signal from inflationay non-Gaussianity. Should a non-Gaussianity of the GR-correction type (and  predicted amplitude) be  detected it would mean that  {\it a)}  initial conditions are of inflationary-type  and  {\it b)} strictly primordial non-Gaussianity is  sub-dominant  to this contribution (e.g., $f_{\rm NL}^{\rm loc}<1.6$).

Moreover, the big difference in the scale-dependent biasing factor between the classes (equilateral, local and enfolded)  of non-Gaussian models implies  that the large-scale halo bias  is  a very sensitive  tool to probe  the shape of non-Gaussianity, highly complementary to other approaches.\\

In Table \ref{tab:1} we show a comparison of  the forecasted constraints on different type of non-Gaussianity  for a selection of planned experiments for CMB  bispectrum and large-scale halo bias. 
\begin{deluxetable}{l  c  c c c c} 
\tablecolumns{6}
\tablewidth{0pc} 
 \tablecaption{Forecasted non-Gaussianity constraints: a) \cite{Yadav} b) \cite{CVM08} c) \cite{cmbpol,Sefusatti} e) This work f) e.g., \cite{Mangilli09} \label{tab:1}}
 \tablehead{ 
\colhead{}   & \multicolumn{2}{c}{CMB Bispectrum} & \colhead{}& \multicolumn{2}{c}{Halo bias} \\ 
\cline{1-6}\\
\colhead{type NG} & \colhead{Planck} & \colhead{(CM)BPol} & &\colhead{Euclid} & \colhead{LSST}}
\startdata
\cutinhead{\hspace{1.5cm}$1-\sigma$ errors}
\rm{Local}& $3^{A)}$ & $2^{A)}$ && $1.5^{B)}$ & $0.7^{B)}$\\
\rm{Equilateral} & $25^{C)}$ &$14^{C)}$&& $-$ &$-$\\
\rm{Enfolded} & ${\cal O}10$ & ${\cal O} 10$&& $39^{E)}$ & $18^{E)}$\\
\cutinhead{\hspace{1.5cm}\#$\sigma$ Detection}
\rm{GR} & N/A & N/A && $1^{E)}$ & $2^{E)}$\\
\rm{Secondaries} & $3^{F)}$ & $5^{F)}$&& N/A & N/A\\
\enddata 
\end{deluxetable} 

This table highlights the complementarity of the two approaches. While forecasted constraints for enfolded non-Gaussianity from CMB data are not available, we can estimate that the errors would be in-between the equilateral and the local case.  One could thus envision different scenarios.

If non-Gaussianity is local with negative $f_{\rm NL}$  and  CMB obtains a detection, then the  halo bias approach should also give a high-significance detection (GR correction and primordial contributions add up), while if it is local  but with positive $f_{\rm NL}$, the halo-bias approach could give a lower statistical significance for small $f_{\rm NL}$ as the GR correction contribution has the opposite sign.

If CMB detects $f_{\rm NL}$ at the level of $\sim 10$ and of a form that is close to local, but halo bias does not detect it, then the CMB bispectrum is given by secondary effects.

If CMB detects non-Gaussianity but is not of the local type, then  halo bias can help discriminate between equilateral and enfolded shapes: if halo bias  sees a signal, it indicates the enfolded type, and if halo bias does not see a signal,  it indicates the equilateral type. Thus even a non-detection of the halo-bias effect, in combination with CMB constraints, can have an important discriminative power.
 
 In any case, if the simplest inflationary scenario holds,  for surveys like Euclid and LSST,   the halo-bias approach is expected to detect a non-Gaussian signal  very similar to the local type signal with an amplitude of $f_{\rm NL}\sim -1.5$ which is  due to large-scales GR corrections  to the Poisson equation. This effect should leave no imprint in the CMB: once again the combination of the two observable can help enormously to discriminate among models for the origin of cosmological structures.

\section*{Acknowledgments}
We  thank N. Bartolo and M. Liguori for useful discussions. 
LV is supported by FP7-PEOPLE-2007-4-3-IRG n. 202182  and MICINN grant AYA2008-03531. 
SM acknowledges  ASI contract I/016/07/0 ``COFIS".

\end{document}